\documentclass[traditabstract]{aa}
\usepackage{graphicx}
\usepackage[varg]{txfonts}
\usepackage{natbib}
\bibpunct{(}{)}{;}{a}{}{,}
\usepackage{array}
\usepackage{xspace}
\usepackage{lscape}
\usepackage{url}
\usepackage{arydshln}
\usepackage{amsmath}
\usepackage{color}
\usepackage[hyperfootnotes=false, linktocpage=true, breaklinks=true, colorlinks=true, linkcolor=blue, citecolor=blue, urlcolor=blue]{hyperref}
\usepackage[all]{hypcap}

\newcommand{\kms}{\ensuremath{\mathrm{km\ s^{-1}}}\xspace}
\newcommand{\cma}{\ensuremath{\mathrm{cm^{-2}}}\xspace}

\newcommand{\NH}{\ensuremath{N_{\mathrm{H}}}\xspace}

\newcommand{\ovii}{\ion{O}{vii}\xspace}
\newcommand{\oviii}{\ion{O}{viii}\xspace}
\newcommand{\neix}{\ion{Ne}{ix}\xspace}
\newcommand{\nvii}{\ion{N}{vii}\xspace}

\mathchardef\mhyphen="2D
\begin{document}

\title{Modelling the disk atmosphere of the low mass X-ray binary \\ EXO 0748-676}

\subtitle{}

\author{
I. Psaradaki \inst{1}
\and
E. Costantini \inst{1}
\and 
M. Mehdipour \inst{1}
\and 
M. D\'iaz Trigo \inst{2}
}

\institute{ SRON Netherlands Institute for Space Research, Sorbonnelaan 2, 3584 CA Utrecht, the Netherlands
                \and
                 ESO, Karl-Schwarzschild-Strasse 2, D-85748 Garching bei Munchen, Germany}

\date{Received 1 August 2018 / Accepted 13 September 2018}

\abstract
{Low mass X-ray binaries exhibit ionized emission from an extended disk atmosphere that surrounds the accretion disk. However, its nature and geometry is still unclear. In this work we present a spectral analysis of the extended atmosphere of EXO 0748-676 using high-resolution spectra from archival XMM-$\textit{Newton}$ observations. We model the RGS spectrum that is obtained during the eclipses. This enables us to model the emission lines that come only from the extended atmosphere of the source, and study its physical structure and properties. The RGS spectrum reveals a series of emission lines consistent with transitions of $\oviii ,\ovii, \neix$ and $\nvii$. We perform both Gaussian line fitting and photoionization modelling. Our results suggest that there are two photoionization gas components, out of pressure equilibrium with respect to each other. One with ionization parameter of $\rm log \ \xi \sim 2.5$ and a large opening angle, and one with $ \rm log \ \xi \sim1.3$. The second component is possibly covering a smaller fraction of the source. From the density diagnostics of the $\ovii$ triplet using photoionization modelling, we detect a rather high density plasma of $>10^{13}$  $\rm cm^{-3}$ for the lower ionization component. This latter component also displays an inflow velocity. We propose a scenario where the high ionization component constitutes an extended upper atmosphere of the accretion disk. The lower ionization component may instead be a clumpy gas created from the impact of the accretion stream with the disk.}

\authorrunning{I. Psaradaki et al.}
\titlerunning{Modelling the disk atmosphere of the LMXB EXO 0748-676}
\maketitle

\section{Introduction}
\par

Low mass X-ray binaries display remarkable physical phenomena such as accretion, winds, plasma heating, time variability and magnetic fields. An accretion disk of an X-ray irradiating plasma encircles the central compact source of the system. Some of these systems present a thickened accretion disk with material that extends above the disk midplane, the so-called accretion disk corona (ADC or hereafter extended disk atmosphere). The nature and exact geometry of this extended atmosphere is not yet fully understood.\par
To this end, different theoretical models predict the existence of a disk atmosphere. \citet{holt1982} studied the accretion disk corona and predicted that it is probably generated by evaporation of hot material from the surface of the accretion disc. Later on, \citet{miller2000} with magnetohydrodynamic (MHD) models showed that an initially weak magnetic field in the core of the disk can be amplified by MHD turbulence driven by magnetorotational instabilities. This field rises out of the disk through buoyancy, creating a magnetised heated corona above the disk at 2-5 scale heights and a corona temperature of $\sim 10^{8}$ K. \citet{garate2002} modelled both the accretion disk atmosphere and the corona, photoionized by a central X-ray source. They found that the vertical scale height of the accretion disk atmosphere is enlarged by illumination heating. In this case the atmosphere is orders of magnitude less dense than the disk itself. \par

Observational studies have also revealed the existence of an extended atmosphere for a number of sources. Using high-resolution X-ray spectroscopy it is possible to identify emission and absorption lines within the disk plasma and around it. \citet{cottam2001_1} studied the XMM-$\textit{Newton}$ RGS spectrum of EXO 0748-676 and found emission and absorption features ($\oviii$, $\ovii$, $\neix$, $\nvii$) from an extended, oblate structure above the accretion disk. 
Further, \citet{garate2003} examined the $\textit{Chandra}$ spectrum of the same source, and found photoionized plasma located above the disk midplane. Also insights have been gained from the spectrum of other accretion disk corona (ADC) sources. 4U 1822-37 \citep{cottam2001_2}, 2S 0921-63 \citep{kallman2003} and Hercules X-1 \citep{garate2005} have been studied using $\textit{Chandra}$ observations. For all these sources spectral signatures of an extended disk atmosphere were found. Noticeably, all the systems above are at high inclination angles. \par
 
Furthermore, the light curve of the low mass X-ray binaries can reveal events such as dips, bursts and eclipses. Dips and eclipses are shown in the light curve as an abrupt drop in the count-rate (see Fig. \ref{fig:lc}). Eclipses are caused when the companion star is passing in front of the compact object and hides the emission of the disk, while the dips can be created due to over-densities from the impact of the accretion stream and the disk \citep{garate2002}. It was found that only a group of low mass X-ray binaries with high-inclination (\textit{i} $\sim 70^{\circ}-90^{\circ}$) present dips and eclipses (\citealt{king1987},  \citealt{trigo2006}). Among the 13 listed LMXBs with high-inclination, only a few of them show eclipses, e.g. EXO 0748-676 \citep{parmar1986} and MXB 1659-298 \citep{sidoli2001}.  \par

EXO 0748-676 is a LMXB discovered in 1985 by the EXOSAT satellite. This source was used as a calibration source of XMM-$\textit{Newton}$ satellite and it has been observed and studied extensively by different satellites (\citealt{parmar1991}, {\citealt{hertz1995}}, \citealt{hertz1997}, \citealt{thomas1997}, \citealt{church1998}, \citealt{garate2003}, \citealt{sidoli2005}, \citealt{wolff2005}, \citealt{peet2017}). \par
 \citet{parmar1986} discovered periodic intensity dips. Eclipses of this source display a period of 3.82 hours lasting 8.3 minutes. The mass of the companion is reported to be $0.08M_{\odot}<M_{c}<0.45M_{\odot}$ and the inclination of the system is $75^{\circ}<i<83^{\circ}$ \citep{parmar1986}. The radial velocity is 20 $\kms$ \citep{duflot1995}. The distance of the source was derived to be [$5.8\pm0.9 $ kpc] or  [$7.7\pm0.9$ kpc] depending on whether the X-ray bursts of the source are hydrogen dominated or helium dominated respectively \citep{wolff2005}. Here, we use the average value of 6.8 kpc.\par
 
In this work, we have analysed the XMM-$\textit{Newton}$ data of EXO 0748-676 during the eclipses. This is the first time that the spectrum from only the extended atmosphere of the disk is being studied for this source. In that way we can understand the emission that comes from the upper disk atmosphere and also its geometry. In previous studies, the density of the plasma has been derived confronting line ratios of the $\ovii$  triplet with theoretical calculations \citep[see][]{porquet2000}. In our study we constrain the density directly using photoionization modelling in SPEX as described in Section \ref{modeling}. This paper is organised as follows. In Section \ref{datared} we present the data and the XMM-$\textit{Newton}$ data reduction and we explain the methodology used to obtain the RGS spectrum of the eclipses. In Section \ref{modeling} we present the Gaussian line fitting and the photoionization modelling of the eclipsed spectrum. Finally in Section \ref{discussion} we discuss the results, proposing a geometry for the X-ray emitting gas.

\section{XMM-\textit{Newton} data reduction}
\label{datared}
\par
 We use data from EPIC-pn \citep{struder2001}, EPIC-MOS \citep{turner2001} cameras and RGS spectrometer \citep{herder2001}, taken from the XMM-$\textit{Newton}$ public archive\footnote{http://nxsa.esac.esa.int/nxsa-web/}. We reduce the data using the Science Analysis Software, SAS (ver. 16). First, we filter the EPIC event lists for flaring particle background. We exclude the observations with flaring particle background exceeding 0.4 counts/s for pn and 0.35 for MOS. 
 \par
\begin{figure*} 
  \centering
  \includegraphics[width=1\textwidth]{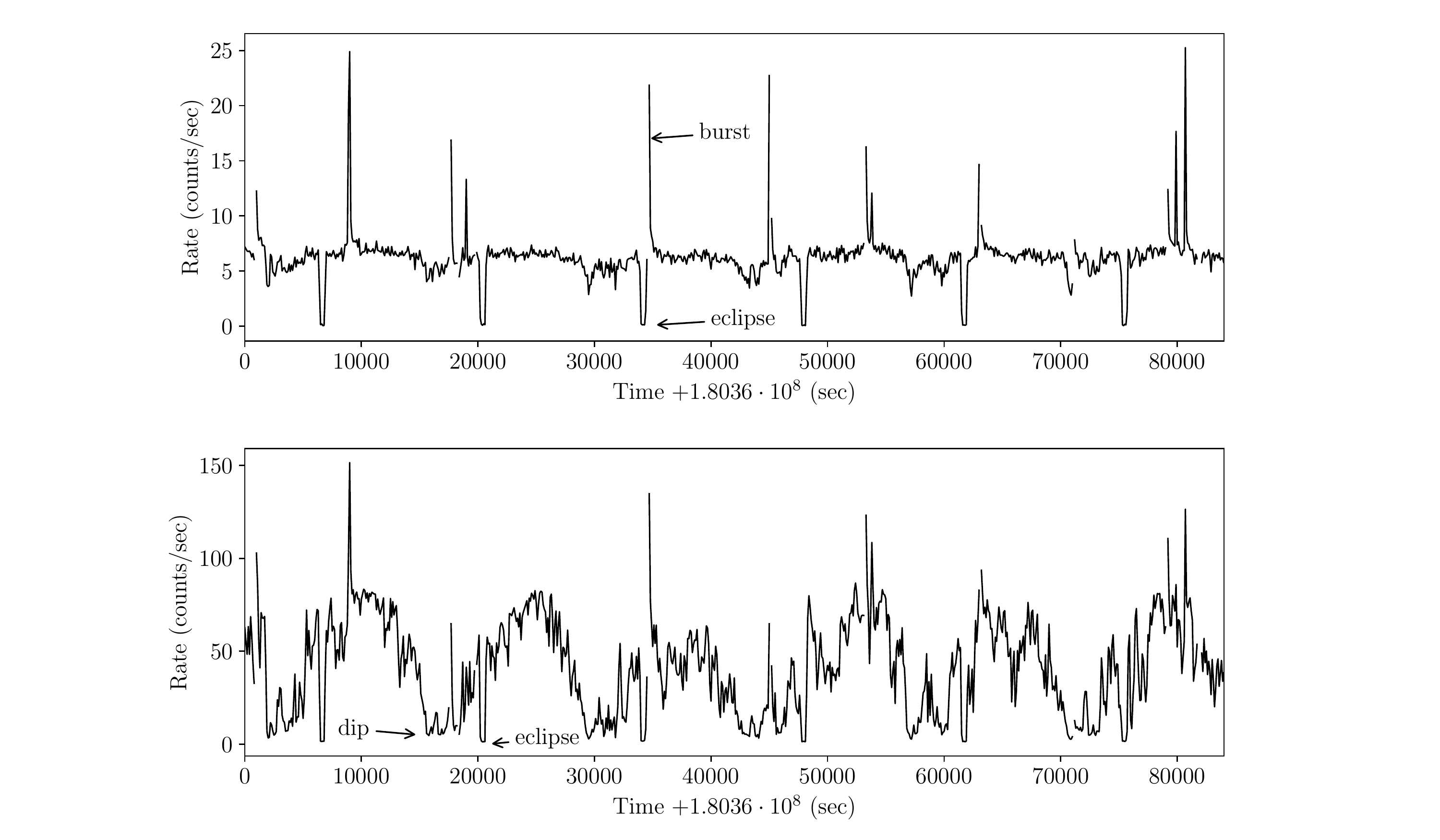}
    \caption{$\textit{Upper panel)}$ Light curve of a single observation showing the eclipses and bursts in the hard band (5-10 keV). $\textit{Lower panel)}$ Light curve for the same observation including dips, eclipses and bursts in the soft band (0.3-5 keV). (Obs. ID: 0160760201) }
\label{fig:lc}
\end{figure*}

 To extract the light-curve we use a circular aperture (R $\sim$ $30^{''}$) around the source. For the background we use a same size circular aperture away from the source but within the same chip. We selected energies between 5 and 10 keV. This allowed us to clearly identify the eclipse events. At softer energies the contribution of the dipping events to the light curve become relevant, possibly confusing the selection. The light-curve of the source is corrected for various effects on the detection efficiency such as vignetting, quantum efficiency and bad pixels, using the SAS task \textit{epiclccorr}. An example of a corrected light curve is shown in Fig. \ref{fig:lc}. In the upper panel we see the light curve of the eclipses and bursts with a 5 < E (keV) < 10 energy selection. In the lower panel we present also the dipping events, obtained in the energy range 0.3-5 keV, and will be discussed in Section \ref{atm}. We call the periods outside these events "persistent emission".  
 \par
We use in total 11 observations obtained in 3 different years, presented in Table \ref{tab:data}. Depending on the availability of the data and the quality of the light curve, we use the EPIC-pn or EPIC-MOS data to obtain the Good Time Intervals (GTI) for the emission during the eclipses. We use these GTIs to extract the RGS spectrum at the time of the eclipses. We choose the best observations according to the quality of the light curves: the count rate (counts/sec) of the persistent emission in the light curve should be at least 2 times higher than the eclipses. After that, we process the RGS data using the SAS task \textit{rgsproc}. We also use a geometrical binning of a factor of 3, which provides a bin size of about 1/3 of the RGS resolution. 
\par
The eclipsed RGS spectra from different years do not present significant variability within the errors. Therefore, we combine the observations using the SAS task \textit{rgscombine} to obtain a better signal-to-noise ratio. We also use a single EPIC-pn observation to create the spectrum of the pn persistent emission and to obtain the continuum parameters. This is useful in order to obtain the correct Spectral Energy Distribution (SED) of the illuminating source which will be discussed in Section \ref{modeling}. 


\begin{table*}[htbp]
\begin{minipage}[t]{\hsize}
\setlength{\extrarowheight}{3pt}
\centering
\caption{Log of XMM-$\textit{Newton}$ observations used in this paper.}
\newcommand{\head}[1]{\textnormal{\textbfh{#1}}}
\begin{tabular}{cccc}
  \hline
  \hline
  obs. ID &instruments& obs. date (year)& observing time (sec)\\
  \hline
  0123500101 &  mos/rgs & 2000 &  62068 \\
 \hline
  0134561101 &  mos/rgs & 2001 &  8312 \\
  0134562101 &  mos/rgs & 2001 &  8678 \\
  0134562401 &  mos/rgs & 2001 &  6875 \\
 0134562501 &  mos/rgs & 2001 &  6882 \\
 \hline
 0160760101 &  pn/rgs & 2003 &  99246 \\
 0160760201 & pn/rgs & 2003 &  98950 \\
 0160760301 & pn/rgs & 2003 &  108703 \\
 0160760401 &  pn/rgs & 2003 &  83951 \\
 0160760601 &  pn/rgs & 2003 &  55351 \\
 0160760801 &  pn/rgs & 2003 &  69750 \\
\hline
\label{tab:data}
\end{tabular}
\end{minipage}
\end{table*}
 

\section{Modelling the eclipsed spectrum}
\label{modeling}
We fit the combined RGS spectrum using the SPEX fitting package (\citealt{kaastra1996}, ver. 3.04). We perform a time averaged analysis of the spectrum. In Figure \ref{fig:gaus} we present the eclipsed spectrum of the source obtained as described in Section \ref{datared}. The observed continuum is extremely low due to the fact that the persistent continuum from the disk and the neutron star is blocked during the eclipse. From the spectrum we can clearly see interesting features such as the emission lines in $\oviii$ (18.97 \AA) and the $\ovii$ triplet (21.6-22.1 {\AA}). For a detailed study of the spectrum we perform a Gaussian line detection and then we model the eclipsed spectrum of the source using a global modelling. \par

\subsection{Line detection}

First, we apply an RGS line detection in the 7-37 {\AA} wavelength range following the method described in \citet{pinto2016}. This will help us to identify fainter lines in our low flux spectrum. We scan the spectrum using Gaussians (component $\textit{gaus}$ in SPEX) in order to identify the lines with the highest significance. We apply a scanning step of 0.025 {\AA} and fix the width of the Gaussian line at 0.005 {\AA}. In Figure \ref{fig:scan} we present the result from the line scanning, where the $\rm \Delta C_{stat}$ is multiplied by the sign of the Gaussian normalization. The red line indicates the level of 3 $\sigma$ detection. From the scan we can clearly see the lines at 18.97 {\AA} and at (21.6-22.1) {\AA} which correspond to $\oviii$ and $\ovii$ respectively and present the highest significance. Also, with lower significance level (but $> 3\sigma$) we observe the $\neix$ line at 13.4 {\AA} and $\nvii$ at 24.7 {\AA}.

\subsection{Gaussian line fitting}
\begin{figure*} 
\centering
	\includegraphics[width=1\textwidth]{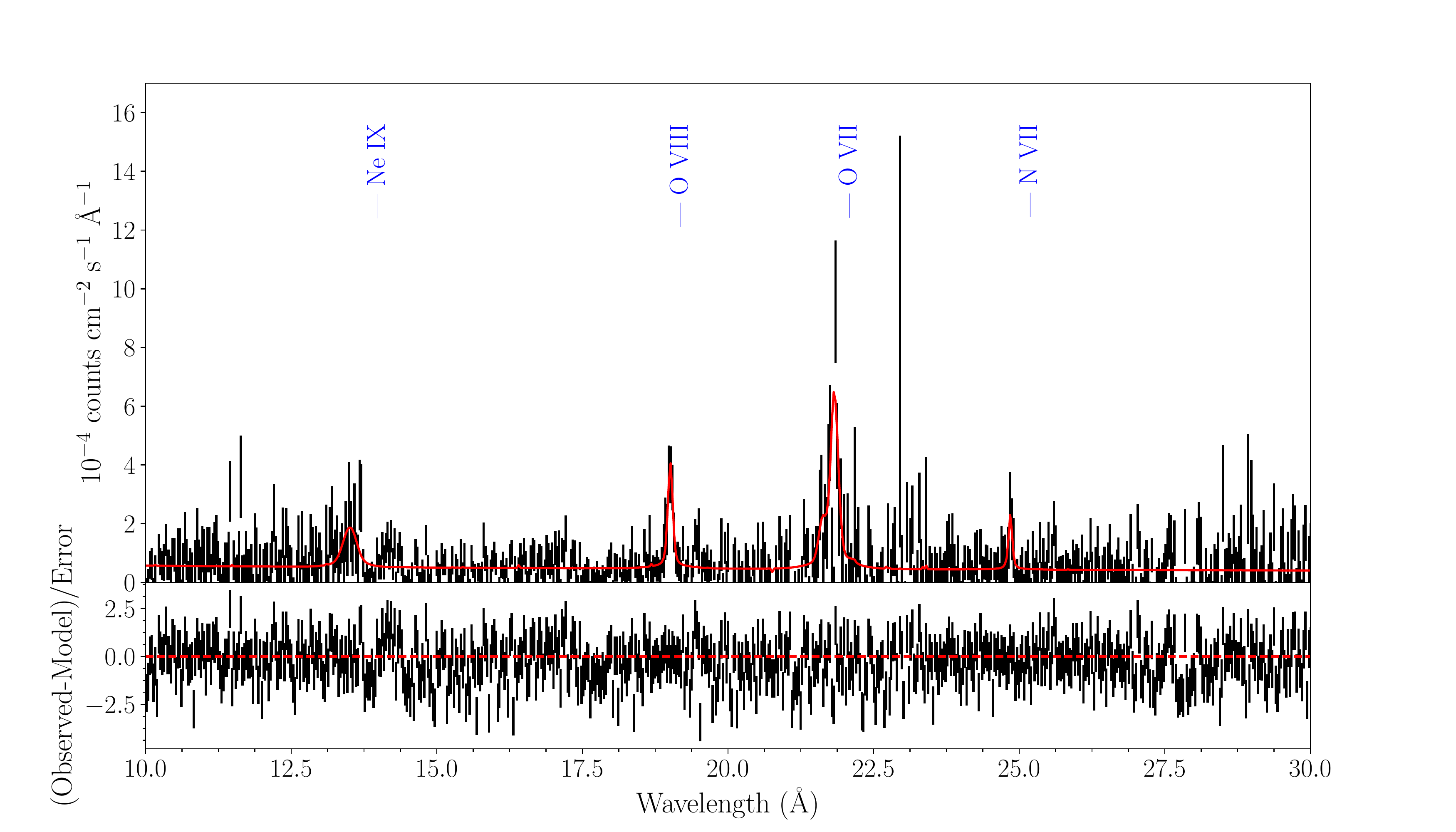}
	\caption{Eclipsed spectrum of EXO 0748-676 with Gaussian modelling and residuals. The line at 23 $\AA$ corresponds to a bad pixel.}
	\label{fig:gaus}
\end{figure*}

\begin{figure} [htbp]
\centering
	\includegraphics[width=0.5\textwidth]{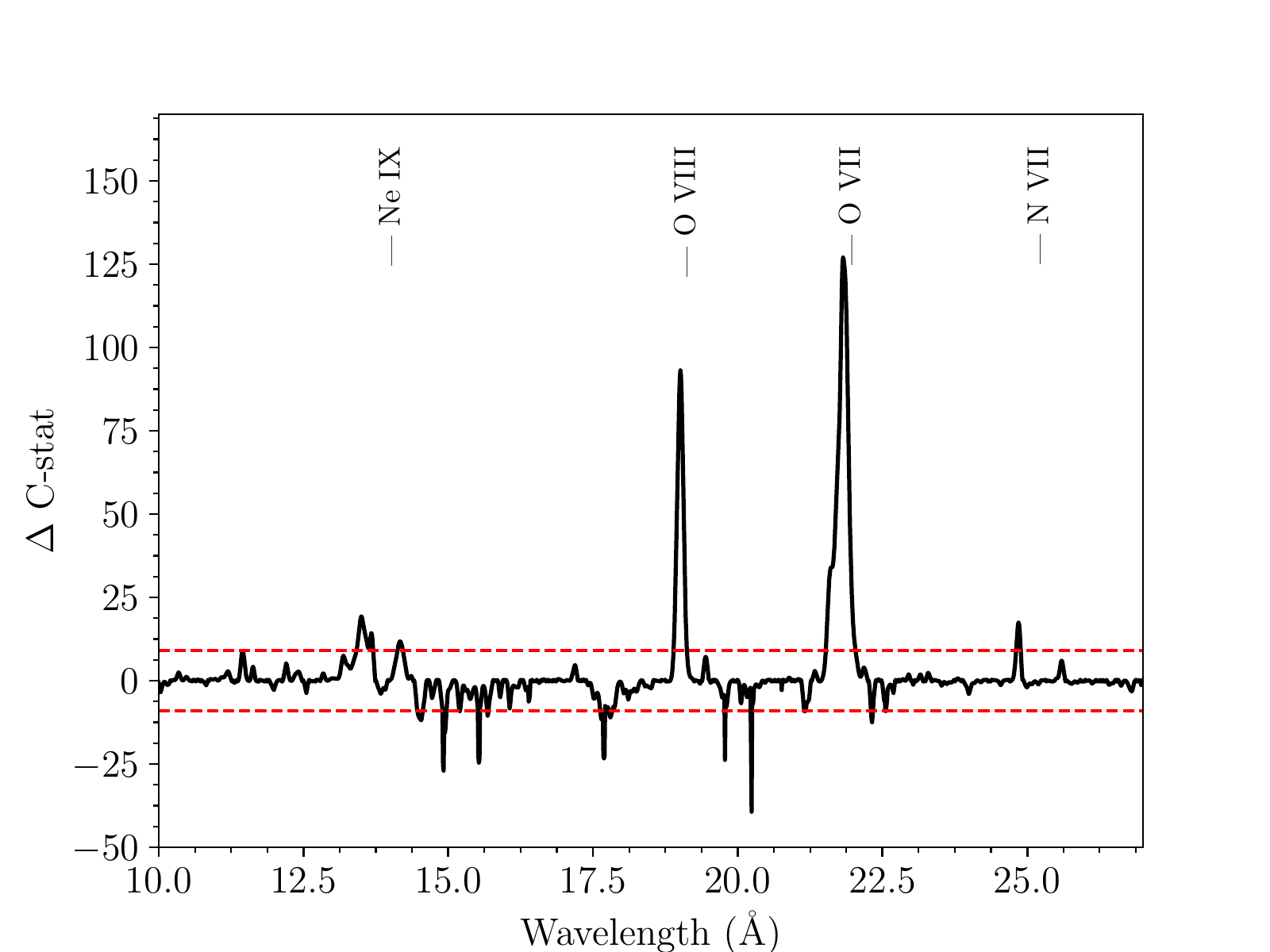}
	\caption{Gaussian line scanning to detect the strongest lines. The red dashed lines correspond to the 3 $\sigma$ detection level. The very narrow absorption-like features correspond to bad pixels.}
	\label{fig:scan}
\end{figure}

We now apply a Gaussian line fitting to parametrise our emission lines in a model independent way and determine the kinematics of our spectrum. We begin with the simple approach of a power law continuum and Gaussian line profiles. In this way we can obtain the strength of the lines and their velocities. We use a power law continuum model for the continuum and the $\textit{hot}$ model in SPEX in order to take into account the Galactic absorption. For the column density we use the Galactic value $3.5 \cdot 10^{21}$ $\cma$ \citep{kalberla2005}. In Table \ref{tab:gaussians} we present the results of the fitting and in Fig. \ref{fig:gaus} the best fit. \par
The width of the lines in some cases could not be resolved. For this reason we couple the widths of the forbidden and the recombination lines to the width of the intercombination line for each of our triplets, $\ovii$ and $\neix$. Also we keep the theoretical values for the wavelength in most of our lines but we apply a redshift component in SPEX for these lines to take into account a possible line shift. We apply different redshift for the $\ovii$ and $\neix$ triplets. We observe our lines slightly redshifted. In Table \ref{tab:gaussians} we present the velocity shift ($v_{flow}$) and broadening ($\sigma$) for each line. The velocity shifts of the lines are comparable. Especially the low ionization ($\ovii$) and high ionization ($\oviii$) lines present the same kinematics. Within the errors, a moderate inflow is detected. \par
From the flux of the lines in the $\ovii$ triplet, we calculate the G and R ratios. These ratios are plasma diagnostics and can be used to identify the photoionized plasma (\citealt{liedahl1999}, \citealt{mewe1999}). The G ratio is sensitive to the electronic temperature while the R ratio is sensitive to the electronic density \citep{gabriel1969} and are given by:

\begin{align}
 & & G  = \frac{f+i}{r}     \\
& & R = \frac{f}{i}         
\end{align}

where $\it f$, $\it i$ and $\it r$ represent the flux of the forbidden, intercombination and resonance line, respectively. The G ratio is $\sim$ 4 which indicates photoionized gas \citep{porquet2000}. The R ratio is $\sim$ 0.02 indicating a high electron density $> 10^{12}$ $ \rm cm^{-3}$ (Fig. 8 of Porquet et al. 2000).


\begin{table*}[htbp]
\begin{minipage}[t]{\hsize}
\setlength{\extrarowheight}{4pt}
\caption{Gaussian modelling parameters of the emission lines detected in the RGS spectrum. The symbol (c) indicates the coupled parameters and (t) the parameters that were set to the theoretical value.}
\centering
\small
\renewcommand{\footnoterule}{}
\begin{tabular}{c  c c c c c c  }
\hline \hline
Line & norm $10^{41}$ (ph/s/keV)  &$\lambda_{observed} (\AA) $&$ \lambda_{theory} (\AA) $ & Width (\AA)  &$\sigma (\kms$)  & $v_{flow}$ ($\kms$)  \\
\hline
$\oviii$  & $2.7^{ +1.7}_{ -0.4}$   & $19.01 \pm 0.01$ & 18.97   &  $0.06 \pm 0.03$ &$ 870 \pm {480}$   & $+584 \pm 126  $ \\
\hline
$\ovii_{r}$   & $2.4 \pm 0.6$  &   21.6 (t)  &  21.6     &  0.13 (c)     & 1750 (c)  &  +300 (c)  \\
$\ovii_{i}$   & $4.2  \pm 0.8$    &  21.8 (t)  &  21.8     &  $0.1^{ + 3.1 \times 10^{-2}}_{ - 2.5\times 10^{-5}} $    & $1750 \pm 400  $  & $ +300 \pm 165 $   \\
$\ovii_{f}$   & $ 4.1^{+ 4.1\times 10^{-5}}_{- 3.8\times 10^{-5}}$    &  22.01 (t)  & 22.01     & 0.13 (c)    & 1750 (c)   &  +300 (c)   \\
\hline
$\neix_{r}$  & $  8.5^{ + 4.4 \times10^{-5}}_{ - 3.7\times 10^{-5}}$ & 13.44 (t)  & 13.44      &   0.2 (c)   &  4400 (c)  & 0 (c)   \\
$\neix_{i}$   & $1.1^{ + 4.2 \times 10^{-4}}_{ - 3.6\times 10^{-4}}$  &  13.55 (t) &  13.55  & $ 0.2^{ + 1.6 \times 10^{-1}}_{ - 5.3 \times 10^{-2}} $  &  $4400^{+ 3500}_{- 1200} $  &$ 0 \pm 2000$ \\
$\neix_{f}$  & $ 1.3 \pm {0.5}$  &  13.69 (t) &   13.69  &  0.2   (c) &  4400 (c)     & 0 (c)   \\
\hline
$\nvii$ & $1.2 \pm 0.3$   & $24.84 \pm 0.01 $ &  24.77    &    0  &  -  & $ +714 \pm{121} $\\
\hline
\label{tab:gaussians}
\end{tabular}
\end{minipage}
\end{table*}



\subsection{Photoionization modelling}

Next, we perform a detailed spectral analysis using a photoionization model. We apply the models for the RGS wavelength range, 7-35 {\AA}.  In order to fit the emission features we use the \textit{pion} model in SPEX\footnote{http://var.sron.nl/SPEX-doc/manualv3.04.00.pdf} (see \citealt{mehdipour2016}). \textit{Pion} is a photoionization plasma model where the photoionization equilibrium is calculated self-consistently.

\subsubsection{Spectral Energy Distribution}
\label{sed}
\begin{figure} [!tbp]
  \centering
  \includegraphics[width=0.55\textwidth]{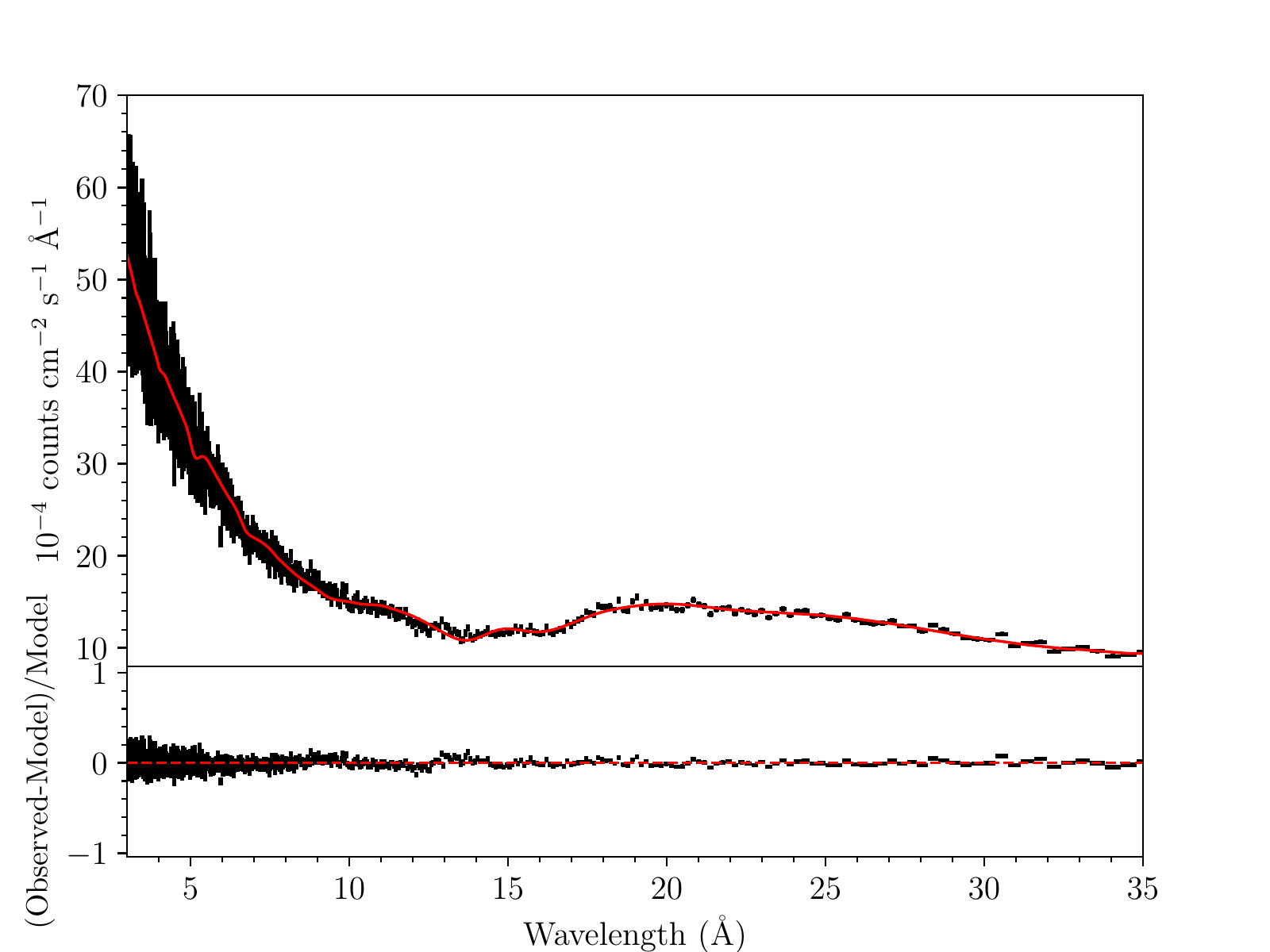}
    \caption{Best fit of the EPIC-pn persistent spectrum and residuals (Obs ID: 0160761301). }
\label{fig:persistent}
\end{figure}

To obtain the correct Spectral Energy Distribution (SED) of the central source we use the following approach. During the eclipses the ionizing continuum is shielded and is not visible in the eclipsed spectrum. Thus, we fit the EPIC-pn spectrum of a single observation, during the persistent phase, to obtain the parameters of the ionizing continuum. The best fit of the persistent emission is shown in Fig. \ref{fig:persistent}. We apply a power-law and a black body. For a better fit of the absorption and emission features we also add a $\textit{hot}$ component for the Galactic absorption. We also use a $\textit{xabs}$ component. $\textit{Xabs}$ is a photoionized absorption model and calculates the transmission of a slab of material. It is needed to fit the ionized gas seen in the persistent spectrum of this source, as described in \citet{peet2017}. $\textit{Xabs}$ is a fast fitting model which contains only absorption lines while \textit{pion} contains also lines in emission. Thus, it is useful to fit the pn spectrum with $\textit{xabs}$ where we cannot see the emission lines because of the moderate resolution of the instrument. The parameters of the fit are listed in Table \ref{tab:persistent}. \par
The emission lines that are present in the 'eclipsed' RGS spectrum are photoionized by the continuum that is seen during the 'persistent' phase. Therefore, for our $ \it pion$ modelling we use the continuum model that we derived from the EPIC-pn spectrum taken during the persistent phase. This continuum is of course different from the low-level observed continuum during the eclipsed phase. Thus, in our SPEX modelling of the eclipsed RGS spectrum, we prevent the persistent continuum from being observed, which is done by incorporating an $\textit{etau}$ model. We note that additional $\textit{etau}$\footnote{http://var.sron.nl/SPEX-doc/manualv3.04.00.pdf} components are used to create the low-energy and high-energy exponential cut-off of the power-law component of the persistent continuum. A high-energy cut-off has applied at 100 keV and a low energy at 13.6 eV.

\subsubsection{Fitting the eclipsed emission}

We fit the following parameters of the \textit{pion} model in SPEX. First, we fit the ionization parameter $\xi =L/ n_{\rm H} \cdot r^{2}$ where $\textit{L}$ is the source luminosity and $ \it r$ the distance from the ionizing source. The hydrogen column density $N_{\rm H}$ in $ \rm cm^{-2}$ and the plasma density $n_{\rm H}$ in $ \rm cm^{-3}$ are also fitted. Further we fit the parameter $\Omega/4\pi$ which gives the opening angle of our plasma material divided by $4\pi$, the velocity shift $v_{flow}$ and the broadening $\sigma$ in $\kms$. \par
\begin{figure*} [htbp]
\begin{centering}
\includegraphics[width=1\textwidth]{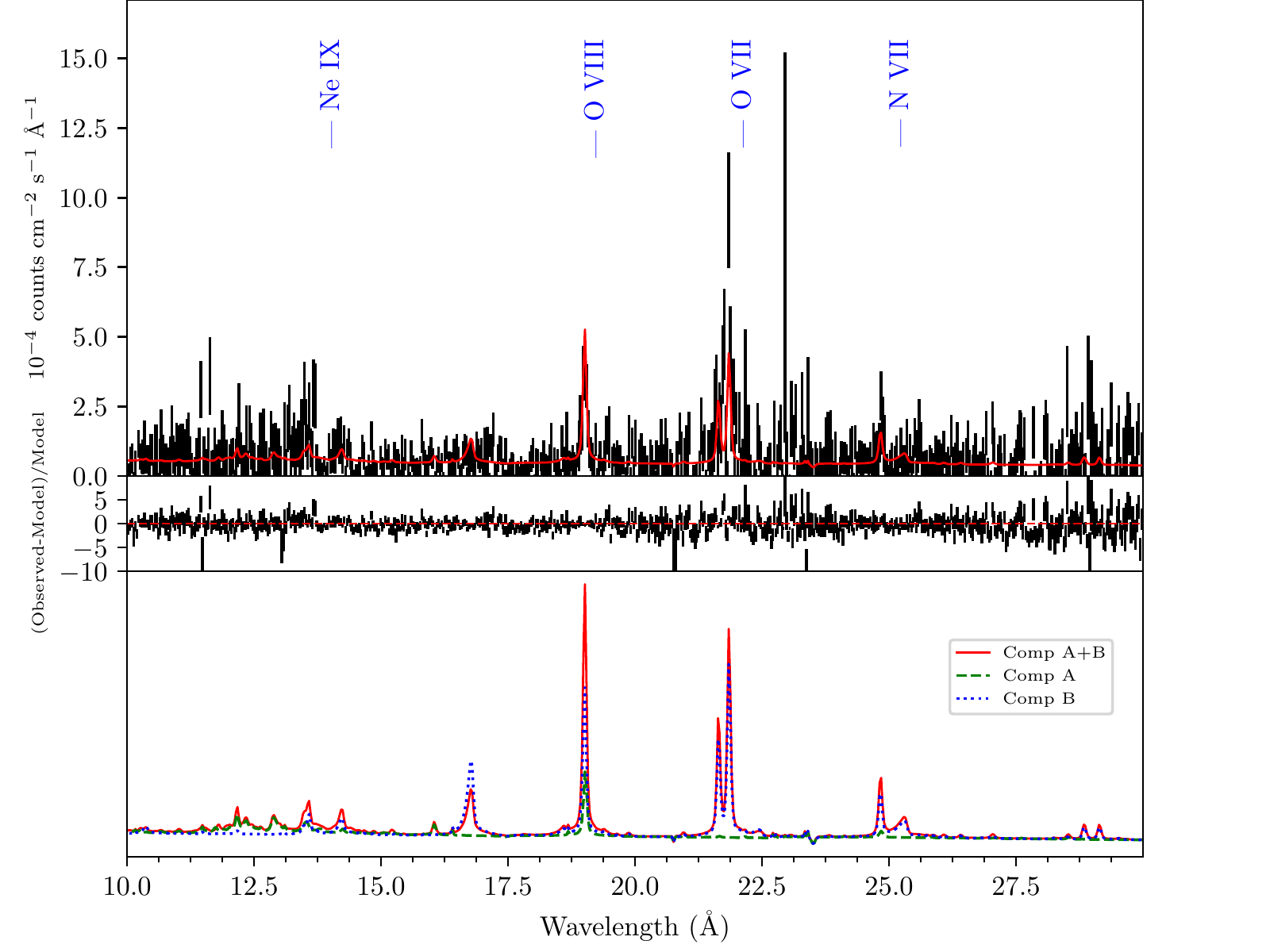}
\label{fig:bestfit}
\caption{$\textit{Upper panel)}$ Best fit spectrum of EXO 0746-676 using 2 pion components. The feature at 23 $\AA$ corresponds to a bad pixel. $\textit{Middle panel)}$ Residuals. $\textit{Lower panel)}$ Pion emission components. The combination of 2 components can fit best both $\ovii$ and $\oviii$.  }
\end{centering}
\end{figure*}

 
The $\oviii$ line prevents the $\ovii$ to be properly fitted when only one photoionization component is applied. For this reason, we need two \textit{pion} components with different ionization parameters to fit the highly ionized and the weakly ionized lines (hereafter, component A and B respectively). A comparison between the modelling with one and two $\textit{pion}$ components is presented in Figure \ref{fig:bestfit}, lower panel. \par 

\begin{figure} [htbp]
\begin{centering}
	\includegraphics[width=0.5\textwidth]{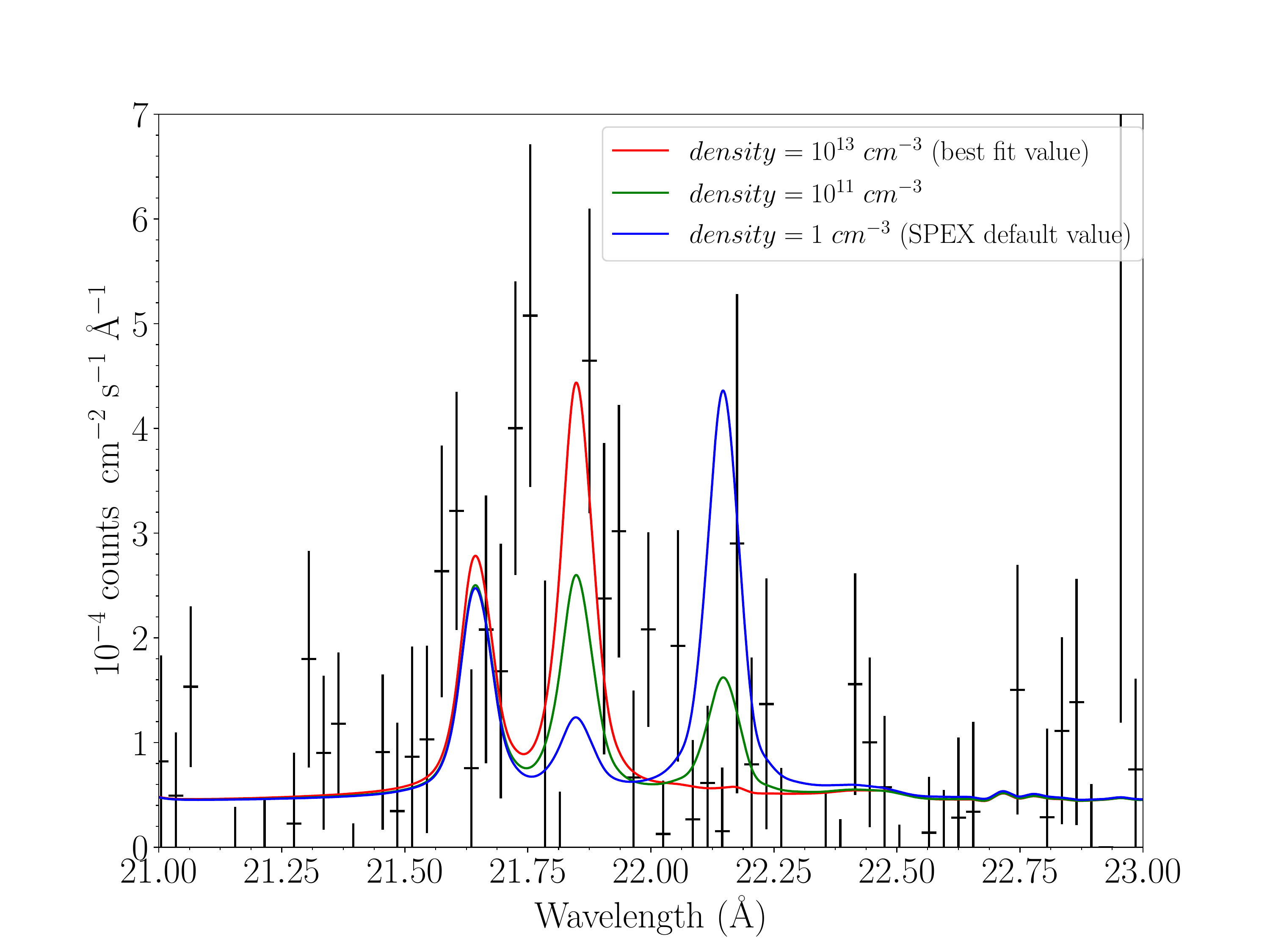}
	\label{fig:oviidensities}
\caption{$\ovii$ spectral range and model calculation with different densities. The best fit is given from the higher density value, displayed in red.}
\end{centering}
\end{figure}

The best fit is shown in Fig. \ref{fig:bestfit} and the best fit parameters for the two photoionization components in Table \ref{tab:pions}. The $\ovii$ triplet (component B) is very sensitive to the density of the plasma. In our model, we applied a density grid starting from a low value of $\sim 1$ $\rm cm^{-3}$ to see the effect of using lower and higher densities. In Figure \ref{fig:oviidensities} we present the zoom in of the $\ovii$ region using the models with different density. The best fit to the data gives a rather high density of  $2 \cdot 10^{13} \rm  cm^{-3}$ for component B (upper limit). Also in our case, the $\ovii$ intercombination line seems to be the strongest and the forbidden line seems to be suppressed which is what we expect for a high density gas \citep{porquet2000}. On the other hand, for component A we cannot constrain a value of density due to poor signal-to-noise around $\neix$ region. The $\oviii$ and $\neix$ lines are not affected from the density changes. For this reason the densities of components A and B are coupled.  \par
 We observe a net gas velocity of $ \sim +800$ $\rm km \ s^{-1}$ (from component B) which indicates an inflowing gas towards the disk. For component A the velocity cannot be constrained. We derive a different ionization parameter and opening angle ($\Omega$) for the two components. In our spectrum, we also detect the $\ovii$ and $\oviii$ radiative recombination emission feature (RRC) at 16.75 {\AA} and 14.20 {\AA} respectively. These narrow features are an indication of photoionized gas (\citealt{paradijs1998}, \citealt{garate2003}).  \par 
Furthermore, knowing the ionization parameter and the density, we calculate the emission measure (EM) for the two photoionization components (see Tables \ref{tab:pions}, \ref{tab:EM}). Knowing the distance from the ionization source from the equation:

\begin{align}
 & & r=\sqrt{\frac{L}{n_H \cdot \xi} }            
\end{align}

we then calculate the emission measure using the formula (see e.g. \citealt{mao2017}):

\begin{align}
 & & EM= n_e \cdot n_H \cdot 4 \cdot \pi \cdot \Omega \cdot  r^{2} \cdot \frac{N_H}{n_H}   
\end{align}

where $n_{H}$, $N_{H}$ and $n_{e}$ are the density, the column density and the electron density, respectively and $\Omega$ is the opening angle. \par

\subsubsection{Testing for alternative models}
\label{cie}

Here, we test whether collisional-ionization equilibrium provides a better description of the data. Therefore, we fitted our eclipsed spectrum with the $\textit{cie}$ model in SPEX which fits the spectrum of a plasma in a collisionally equilibrium state. We use the same continuum (as described in Section \ref{sed}) and fit two $\textit{cie}$ components. We also add a $\textit{reds}$ component in SPEX to take into account the possible redshift of the lines. We found a different plasma temperature for each $\textit{cie}$ component, 5 keV and 0.15 keV respectively. In this model, we notice that the forbidden line is strong and the intercombination is suppressed which does not represent our case. We conclude that a low-density collisionally ionized gas is not valid in this case. We test also a high-density collisionally-ionized gas, which also does not fit better in our spectrum (C-stat\,/\,Exp. C-stat= 1842\,/\,1067). \par
\begin{figure*} [!tbp]
  \centering
  \includegraphics[width=0.95\textwidth]{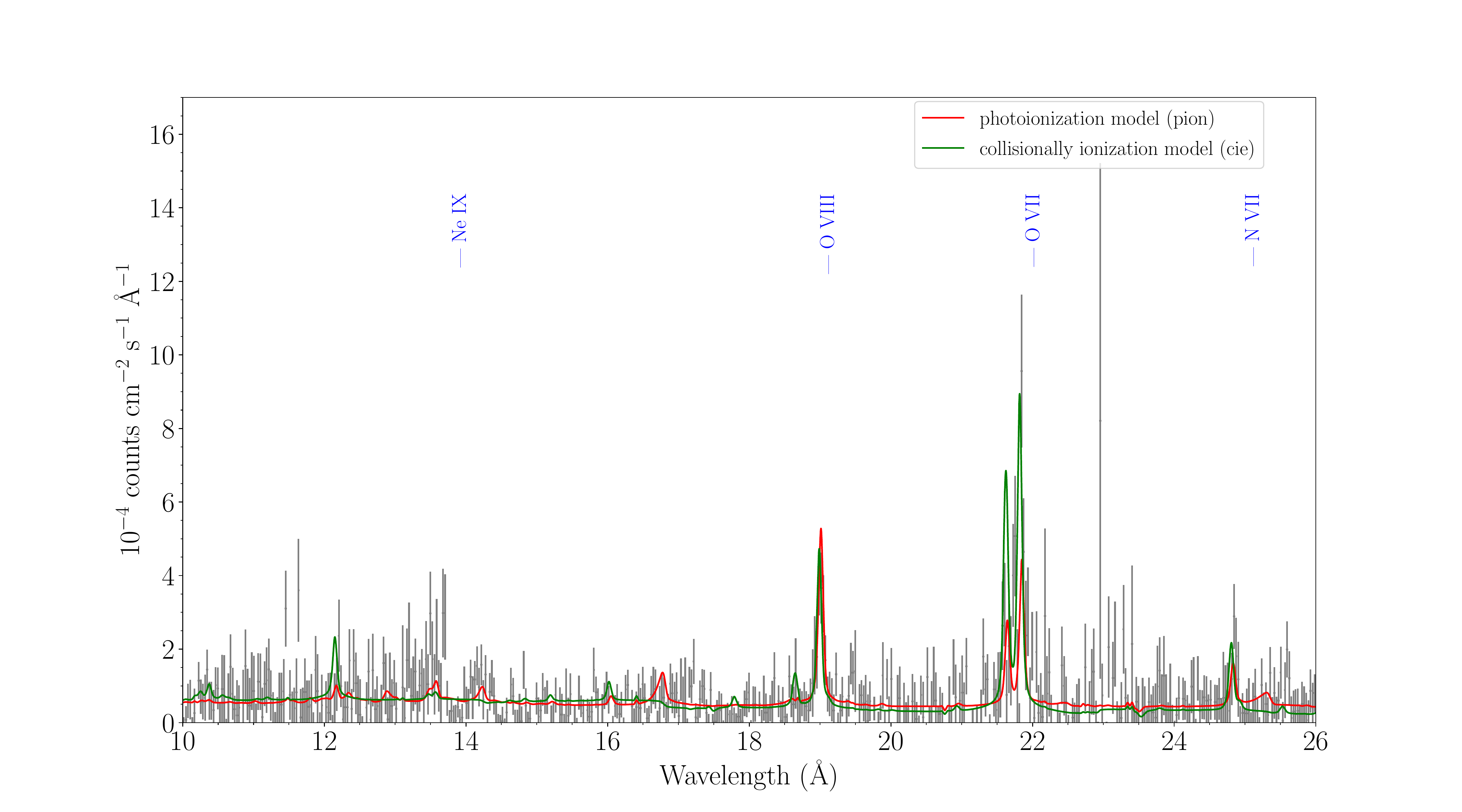}
    \caption{Modelling the eclipsed spectrum of EXO 0748-676 with collisionally ionized gas and comparison to the best-fit photoionized gas. In both cases we have a high-density gas.}
\label{fig:cie}
\end{figure*}

 Furthermore, in the $\textit{cie}$ model, one can test the case of non equilibrium plasma. We let free the parameter $\textit{rt}$ (the ratio of ionization balance to electron temperature). The electron density for both components is $\sim 3 \cdot 10^{13} \rm  cm^{-3}$. In this case, the intercombination line becomes stronger than the resonance and the forbidden is suppressed but we still do not obtain a better fit than the photoionized case, as we see in Fig. \ref{fig:cie} (C-stat\,/\,Exp. C-stat= 1738\,/\,1067). Lastly, we tested the possibility of a more complicated model which is the combination of a collisionally ionized component and a photoionization component in order to reduce our residuals. It still does not improve our fit so we conclude that the best fit model that represents our case is the photoionization modelling.


\begin{table*}[!tbp]
\begin{minipage}[t]{\hsize}
\centering
\setlength{\extrarowheight}{3pt}
\caption{The continuum parameters of the spectrum during the persistent emission and the power law parameters resulting from the eclipsed spectrum. The symbol 'pow' refers to the power law parameters and 'bb' to the black body.}
\small
\renewcommand{\footnoterule}{}
\renewcommand{\arraystretch}{1.3}
\begin{tabular}{c c c c c c }
\hline \hline
Continuum    &   Parameter (Unit)       &  Value    \\
Component   &   &   \\
\hline
$pow_{persistent} $    & norm $10^{44}$ (ph/s/keV at 1 keV)   & $1.6 \pm 0.02 $ \\
       & $\Gamma$ (photon index)   & $1.4 \pm 0.01$   \\
$bb_{persistent} $    &  norm ($\rm cm^{2}$)      & $1.13^{ + 8.0 \times 10 ^{-8}}_{ - 5.5  \times 10^{-8}} $ \\
         & Temperature (keV)    &$ 0.12 ^{+ 1.8 \times 10 ^{-3}}_{- 1.7 \times 10^{-2}}  $ \\
\hline 
$pow_{eclipsed} $   &  norm $10^{44}$ (ph/s/keV at 1 keV)   &$ 0.032 \pm 0.003$\\
         &$ \Gamma$    &$ 1.9 \pm 0.2 $\\

\hline
\label{tab:persistent}
\end{tabular}
\end{minipage}
\end{table*}


\begin{table*}[!tbp]
\begin{minipage}[t]{\hsize}
\setlength{\extrarowheight}{3pt}
\caption{Best-fit parameters of the pion photoionization model components fitted to the stacked XMM-$\textit{Newton}$ data. The symbol (c) indicates the coupled parameters.}
\centering
\small
\renewcommand{\footnoterule}{}
\begin{tabular}{c c c c c c c c c c c }
\hline \hline
Comp      & $\log~\xi$            &$ \NH$           & $\sigma  $            & $\Omega\,/\,4 \pi$ & $n_{H}$ & $v_{flow}$  \\
  & ($\rm erg \cdot \rm s^{-1} \cdot \rm cm$)     & ($ \rm cm^{-2} \cdot 10^{21}$)   & ($\kms$)    &   -    &  $(  \rm cm^{-3} \cdot 10^{13})$ &  ($\kms$)    \\
\hline
A     & $2.5\pm 0.1$     & $0.75^{ + 0.53}_{ - 0.25}$ & $591 \pm 350$     & $< 0.5^{+ 0.0}_{- 0.3}$  & 2 (c) & 0 (fixed)  \\
B     & $1.3 \pm 0.1$   &  $18^{ + 7}_{ - 6}$ & $  504^{+ 190}_{- 120} $ & $0.007\pm0.002 $ & $ < 2^{+ 0.0}_{- 0.6}$  & $880 ^{+28}_{-91} $     \\

\hline
\multicolumn{8}{c}{C-stat\,/\,Exp. C-stat = 1637\,/\,1067 \footnote{For the Exp. C-stat see \citet{kaastra2017}} } \\
\hline

\label{tab:pions}
\end{tabular}
\end{minipage}
\end{table*}



\begin{table*}[!tbp]
\begin{minipage}[t]{\hsize}
\setlength{\extrarowheight}{3pt}
\caption{Calculated parameters according to the best-fit parameters of the pion photoionization model components fitted to the stacked XMM-$\textit{Newton}$ data.}
\centering
\small
\renewcommand{\footnoterule}{}
\begin{tabular}{c c c c c c }
\hline \hline
Comp    & L & EM & r & $n_e$ & Thickness \\
  &  ($ \rm erg \cdot s^{-1}$) & $( \rm cm^{-3})$ & (cm) &$ (\rm 10^{9} cm^{-3} )$ & (cm)   \\
\hline
A      &$1.83\times 10^{32}$& $ 3\times 10^{51} $ & $1.67\times 10^{8} $ & 2.4& $ 3.5 \times 10^7$\\
B      &$7.68\times 10^{30}$& $ 9\times 10^{46}$&$ 2.4 \times 10^{10}$ & 2.4 &$8.2 \times 10^8$    \\

\hline
\label{tab:EM}
\end{tabular}
\end{minipage}
\end{table*}


\subsection{Are the two photoionization components thermally stable?}
\label{scurve}



\begin{figure*} [h]
\centering
  \includegraphics[width=0.5\textwidth]{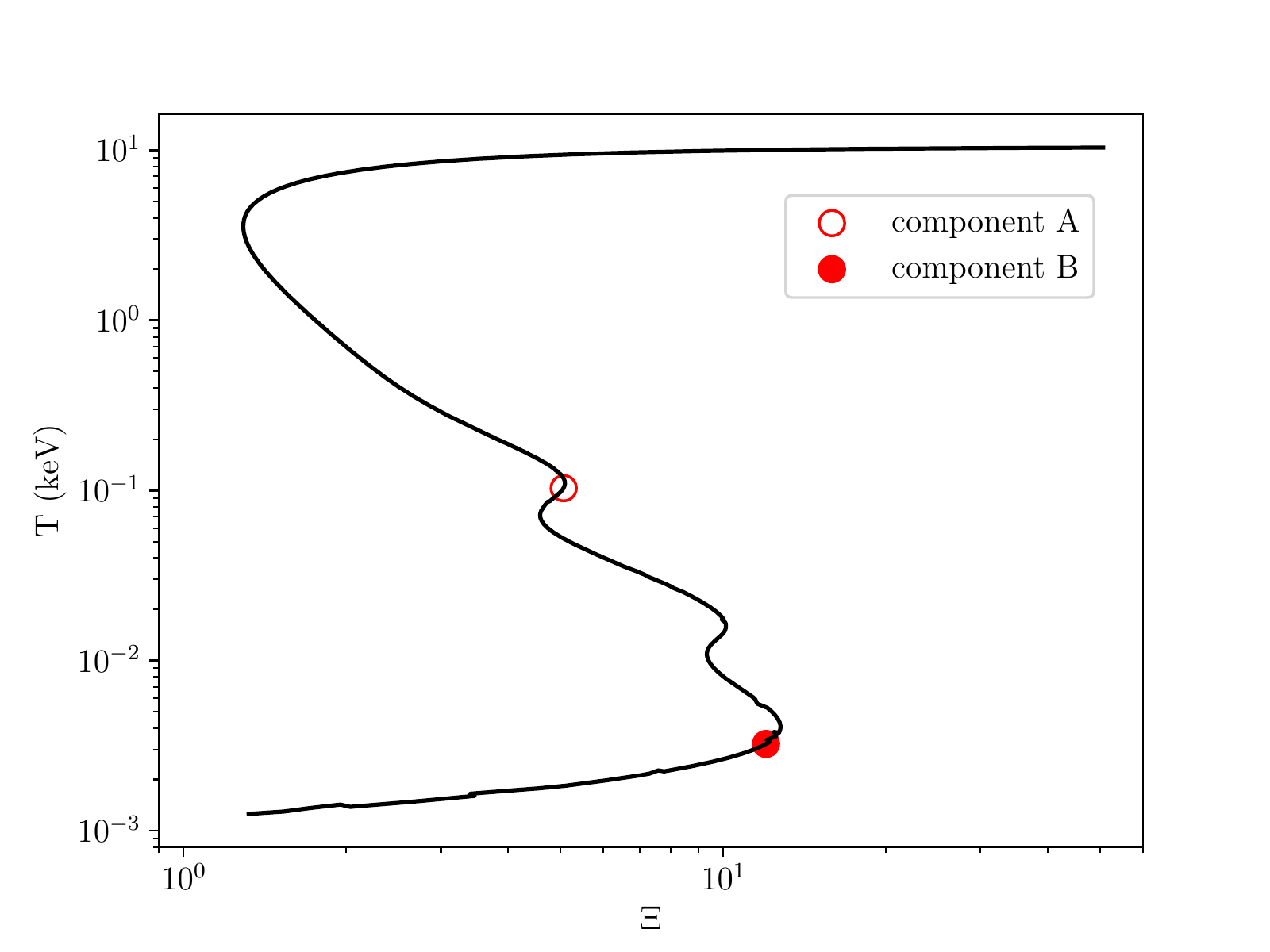}
    \caption{Thermal stability curve of the photoionization modelling. The two components are not in pressure equilibrium. }
\label{fig:scurve}
\end{figure*}


A photoionized plasma can be unstable at some ionization states. We test whether our photoionization components are in pressure equilibrium. This can be tested by creating a stability curve (called also S-curve or cooling curve). The stability curve shows the change in the electron temperature as a function of the pressure-form of the ionization parameter $\Xi$, introduced by \citet{krolik1981}, which is defined as the radiation pressure divided by the gas pressure.
 \par 
We produce the S-curves as follows. We first create a grid of electron temperatures and ionization parameter $\xi$ according to our photoionization modelling and the SED in SPEX. Further we calculate the pressure, $\Xi$. The parameter $\Xi$ gives the ionization equilibrium and can be expressed as $\Xi =F/n_{H}\cdot c\cdot kT$ where $n_{\rm H}$ is the hydrogen density in $ \rm cm^{-3}$, $\it k$ is the Boltzmann constant and  $\it T$ the gas temperature. Since $F=L/4\cdot \pi \cdot r^{2}$ and $\xi=L/n_{H} \cdot r^{2}$, $\Xi$ can be written as:

\begin{equation}
\Xi=\frac{L}{4 \pi  r^{2} \cdot n_{H} \cdot c \cdot kT}=\frac{\xi}{4 \pi \cdot c \cdot kT}=19222 \cdot \frac{\xi}{T}
\end{equation}

 In Figure \ref{fig:scurve} we present the stability curve. We also overplot with empty and filled circle the location of our photoionization components A and B, respectively. The two components are not in pressure equilibrium. Component B is cooler and it belongs to a thermally stable part of the curve. Component A is hotter and it may reside on a thermally unstable branch, although from the position of the component on the curve this is difficult to constrain.  

\section{Discussion}
\label{discussion}

In this work, we performed a spectroscopic study of the eclipsed spectrum of the low mass X-ray binary EXO 0748-676. We fit the spectrum using both Gaussian line modelling and photoionization modelling. By modelling the eclipsed spectrum we isolate only the material that come from the area that surrounds the disk, from which we confirm the existence of an upper extended disk atmosphere. \subsection{Geometry of the system}
\label{geometry}

\subsubsection{The observed two-phase gas}
\label{equil}

In our analysis we performed photoionization modelling using two components in order to fit both the lower ($\ovii$) and higher ionization ($\oviii$) lines (see Figure \ref{fig:bestfit}). The Gaussian line fitting and G ratio indicate the existence of photoionized emission from the upper disc atmosphere. Also, with our global modelling, we found narrow RRC features in the spectrum. The RRC emission features in the spectrum show that photoionization is the dominant ionizing mechanism in the plasma. Thus, we support our initial assumption for using photoionization modelling. \par
From the stability curve (see Section \ref{scurve}) we see that we have a two-phase gas out of pressure equilibrium. This may suggest that components A and B (see Table \ref{tab:pions}) are independent gas components. Looking at the fitting parameters in Table \ref{tab:pions}, the two components have different ionization ($\xi$) and opening angle ($\Omega$). The higher ionization component (A) comes from a region with opening angle of $2 \cdot\pi$ while the lower ionization one (B) comes from a region two orders of magnitudes smaller. We also find a rather high density for component B  ($\sim 10^{13} \rm cm^{-3}$) while for component A the density could not be determined due to poor signal-to-noise around the $\neix$ region. Component B seems to be also inflowing. For component A the velocity cannot be constrained.  \par

\subsubsection{The structure of the atmosphere}
\label{atm}

\begin{figure*} [htbp]
\includegraphics[width=0.95\textwidth]{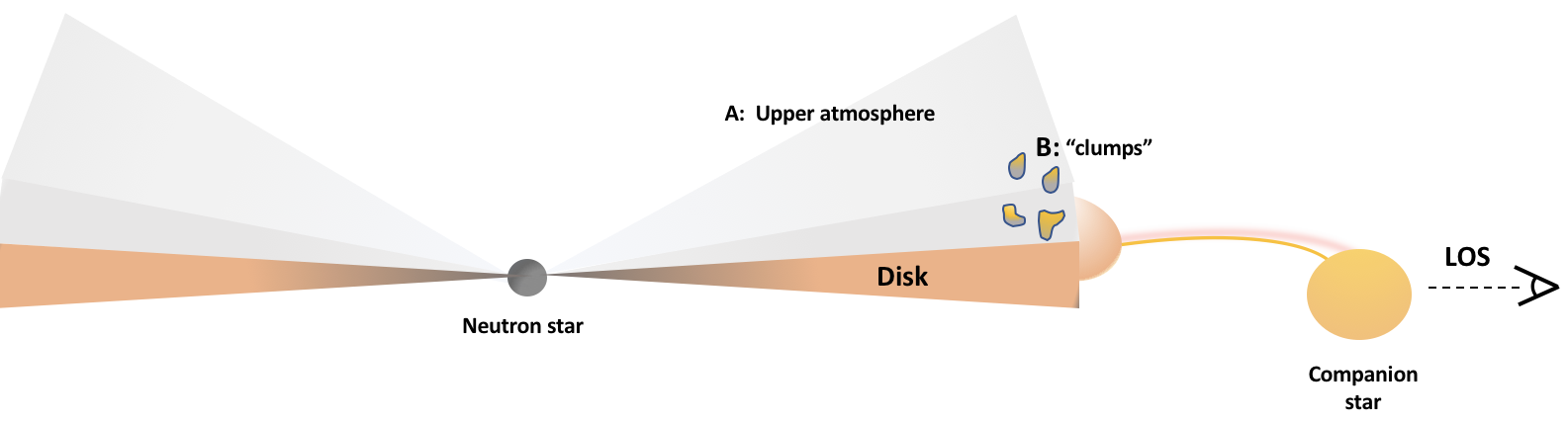}
\caption{Illustration of EXO 0748-676 with the clumps in the extended atmosphere. Symbols A and B show the emission regions of our two photoionization components. A is the hotter component while B is the cooler one. The components in the figure are not to scale.}
\label{fig:exo}
\end{figure*}

In Fig. \ref{fig:exo} we present an illustration that shows the position of the two-phase gas and the shape of the upper atmosphere which most likely fit to our case. From our observational results we see a two-phase gas covering a different area along the line of sight. The emission measure for the lower ionization component (B) is smaller than that of the higher one (A) (see Table \ref{tab:EM}). From the definition of the emission measure ($EM=\int n_{e} n_{H} dV$), we can conclude that the gas of component B is coming from a small volume which is consistent with the small value of $\Omega$, while component A is covering a broader area. \par
 A likely scenario that explains the above results is the following. The high $\xi $ and larger $\Omega$ material comes from the extended atmosphere that is located above the accretion disk. This atmosphere can be created by illumination that comes from the disk which is heated by the central X-ray source (see \citealt{garate2003}). \par 
The material that we see during the eclipse, and refer to component B, can be related to the material impinging into the disk. Interestingly, the eclipses are systematically coming during the dipping event, almost at the end of it (see Fig \ref{fig:lc}, lower panel). This applies to all our observations. Most dips are observed to be over-densities above the disk region (at the outer disk edge) which has been thickened as an effect of the impact with the accretion stream (\citealt{white1982}, \citealt{king1987}). \citealt{trigo2009} found that in low mass X-ray binary system XB 1254-690 , the gas in the line of sight causing the dips is clumpy. Clumps have also been suggested to explain the phenomenology seen in Her X-1 \citep{schandl1996} or to explain the high optical luminosity of supersoft X-ray sources \citep{suleimanov2003}. In our case, the lower $\xi $ gas with the small volume may come from clumpy region above the disk. This emission could come from a clumpy gas which may have been created due to pressure instabilities during the impact of the accretion stream and the disk. The clumps seem to have a rather high density ($\sim 10^{13} \rm cm^{-3}$) and also they might have an inflowing velocity. This geometrical picture can explain also our findings for a two-phase gas.  \par
The atmosphere (component A in Fig. \ref{fig:exo}) is most likely extended in the outer region of the disk. We can explain the existence of an extended corona in the frame of X-ray irradiation of the upper layers of an accretion disk (see e.g. \citealt{garate2002}). In this context, in the inner region of the accretion disk, close to the neutron star, the viscous heating is the dominant heating mechanism. The outer region of the accretion disk is dominated by external illumination and thus radiative heating exceeds viscous heating. The X-ray field of the neutron star photoionizes and heats the gas. Additional heating also is provided by the illumination from the accretion stream. The gas tends to move towards a situation of hydrostatic equilibrium, convection is suppressed and the disk is increasing the scale height. For this reason the geometry of the extended atmosphere is more extended in the outer part of the disk.

    
\subsection{Comparison with previous results}
\label{comparison}

We compare the spectroscopic results of EXO 0748-676 with results from other sources that present an extended accretion disk corona. \citet{kallman2003} analysed $\textit{Chandra}$ observations from the high-inclination ADC source 2S 0921-63. From the density diagnostics in $\ovii$ line, they constrain a density for the ADC of $10^{9}-10^{11} \rm cm^{-3}$ and the ionization parameter, $ \rm log \ \xi=2$.\par

Further, \citet{cottam2001_2} found from the $\textit{Chandra}$ spectrum of the ADC source 4U 1822-37 an opening angle of $\frac{\Omega}{2\pi}=0.11$ which, according to the authors, constrains an extended area around the source. Also they derive an electron density of $10^{11}$ $\rm cm^{-3}$. Further, \citet{garate2005}, for the ADC source Hercules X-1, obtained using the R-ratio a high density of $10^{13} \rm cm^{-3}$, similar to the value obtained here. According to the authors, this density is consistent also with predicted atmospheric models. Also, in both sources, the presence of the RRC emission features is evident.  \par

\citet{cottam2001_1} studied the XMM-$\textit{Newton}$ spectrum during the persistent emission intervals of EXO 0748-67 and found emission lines from $\ovii$, $\oviii$, $\neix$ and $\nvii$. They conclude that these features are coming from the extended atmosphere above the disk. They estimate the density using the line ratios and they found a lower limit for $\ovii$ and $\neix$ of $ 2 \cdot 10^{12} \rm cm^{-3}$ and $ 7 \cdot 10^{12} \rm cm^{-3}$. They also found the velocity broadening of the $\rm Ly\alpha$ of $\oviii$ $\sim 1400$  $\rm km \ s^{-1}$ and a systemic velocity of the emitting plasma < 300 $\rm km \ s^{-1}$. \par

\citet{garate2003} analysed the persistent $\textit{Chandra}$ spectrum of EXO 0748-676 and confirmed the existence of an upper atmosphere around the disk. They find the upper atmosphere to extend $8^{\circ}-15^{\circ}$ with respect to the disk midplane. Further the density measured from the $\ovii$ lines is $\sim 10^{11}$ $\rm cm^{-3}$ and a mean broadening of the brightest lines of $\sim 700$  $\rm km \ s^{-1}$. Their lines seem to be at rest, while our modelling shows a slight redshift. It has to be noted that in this case the authors studied the total emission of the disk in the persistent interval while in our case we get the emission only from the upper disk atmosphere and to a specific time interval of the accretion event. \par

\section{Summary}
\label{Summary}

In this study we have analysed the XMM-\textit{Newton} RGS spectrum of the low mass X-ray binary EXO 0748-676 during the eclipses. This allowed us to study for the first time the gas coming only from the upper disk atmosphere. In our work, we modelled the emission lines using Gaussian line fitting to estimate the gas velocity and line shift in a model independent way. Furthermore, we used photoionization modelling and constrained the density and the structure of the atmosphere. Our conclusions are the following: \\
\begin{itemize}
    \item We confirm the existence of an extended atmosphere above the accretion disk of EXO 0748-676. We detect $\ovii$ and $\oviii$ lines with high significance but also $\neix$ and $\nvii$ lines. We measure positive velocity shifts for the strongest lines which gives an evidence for inflowing gas towards the central source.
    \item From the line ratios of the $\ovii$ triplet and the photoionization modelling, we estimate the density of the gas and we obtain a rather high value of $\sim 10^{13} \rm cm^{-3}$. 
    \item In our modelling we use two photoionization components. Our thermal stability analysis shows that the two components are out of equilibrium with each other. This means that we probably observe two distinct gas components. One displays smaller opening angle and is coming most likely from clumps created from the impact of the accretion stream with the disk, inflowing towards the source. The other belongs to the atmosphere and from our modelling we find that is covering an area of at most $2\pi $. 
     \item The results support the scenario that the extended disk atmosphere is created due to heating of the outer part of the disk from the central compact source and the accretion stream and it is most likely photoionized. 
\end{itemize}

\begin{acknowledgements}
The authors thank the anonymous referee for the useful comments. IP, DR and EC are supported by the Netherlands Organisation
for Scientific Research (NWO) through The Innovational Research Incentives
Scheme Vidi grant 639.042.525. The Space Research Organization of the
Netherlands is supported financially by the NWO. We would like to thank R. Waters for constructive suggestions on the manuscript and C. Done for useful discussions on the shape of the disk atmosphere. 
We also thank I. Urdampilleta for providing help with $\textit{python}$ and C. Pinto for advises on the line detection. \end{acknowledgements}

\vspace{-0.4cm}


\end{document}